\documentclass[aps,prb,groupedaddress,10pt,
,twocolumn
]{revtex4-1} 
\usepackage{graphicx,graphics,amsfonts,amssymb,color,bm,verbatim,ifsym,amsmath}

\def\be{\begin{equation}} 
\def\ee{\end{equation}} 
\def\ba{\begin{eqnarray}} 
\def\ea{\end{eqnarray}} 
\def\bc{\begin{center}} 
\def\ec{\end{center}}

\begin{document} 

\title{Optical Kerr Effect in Graphene: \\ Theoretical Analysis of the Optical Heterodyne Detection Technique
} 

\author{N. A. Savostianova} 
\affiliation{Institute of Physics, University of Augsburg, D-86135 Augsburg, Germany} 

\author{S. A. Mikhailov} 
\email[Electronic mail: ]{sergey.mikhailov@physik.uni-augsburg.de} 
\affiliation{Institute of Physics, University of Augsburg, D-86135 Augsburg, Germany} 
\date{\today} 

\begin{abstract}
Graphene is an atomically thin two-dimensional material demonstrating strong optical nonlinearities including harmonics generation, four-wave mixing, Kerr, and other nonlinear effects. In this paper we theoretically analyze the optical heterodyne detection (OHD) technique of measuring the optical Kerr effect (OKE) in two-dimensional crystals and show how to relate the quantities measured in such experiments with components of the third-order conductivity tensor $\sigma^{(3)}_{\alpha\beta\gamma\delta}(\omega_1,\omega_2,\omega_3)$ of the  two-dimensional crystal. Using results of a recently developed quantum theory of the third-order nonlinear electrodynamic response of graphene, we analyze the frequency, charge carrier density, temperature, and other dependencies of the OHD-OKE response of this material. We compare our results with a recent OHD-OKE experiment in graphene and find good agreement between the theory and experiment. 
\end{abstract} 


 
\maketitle 


\section{Introduction\label{sec:Intro}}

The nonlinear electrodynamics and optics of graphene \cite{Novoselov04,Novoselov05,Zhang05} have evolved into an active field of research in recent years. It was predicted\cite{Mikhailov07e} in 2007 that, due to the linear energy dispersion of graphene quasi-particles, this material should demonstrate a strongly nonlinear electrodynamic response. This prediction was confirmed in a number of experiments, in which the higher harmonics generation \cite{Dragoman10,Bykov12,Kumar13,Hong13,Soavi17}, four-wave mixing\cite{Hendry10,Gu12,KonigOtto17,Alexander17}, saturable absorption \cite{Bao09,Zhang09,Popa10,Popa11,Zheng12,Bianchi17}, the Kerr effect \cite{Zhang12,Chen13,Miao15,Dremetsika16,Vermeulen16,Dremetsika17} and other nonlinear phenomena \cite{Mics15,Sharif16,Sharif16a,Brida13,Tomadin13} in graphene have been observed. Theoretically, the higher harmonics generation \cite{Mikhailov08a, Mikhailov09a,Dean09,Dean10,Avetissian13,Smirnova14,Cheng14b,Wang16,Cheng17, Rostami16a},
nonlinear plasma-wave related effects  \cite{Mikhailov11c,Mikhailov12c,CoxAbajo14,CoxAbajo15,CoxAbajo16, Yao14,Mikhailov17b},
nonlinear cyclotron resonance \cite{Mikhailov09b,Tokman14}, and saturable absorption \cite{MariniAbajo16} have been studied. The influence of the dielectric environment on the harmonics generation from graphene has been discussed in Refs. \cite{Savostianova15,Savostianova17a}, the nonlinear effects in a gapped graphene in Ref. \cite{Jafari12}, a nonlinear time-domain optical response has been considered in Ref. \cite{Ishikawa10}, and other aspects of the nonlinear graphene response have been discussed in Refs. \cite{Mikhailov16b,Khurgin14,Mikhailov17c,Khurgin17}; see also review articles \cite{Glazov14,Hartmann14}. Recently a quantum theory of all third-order nonlinear effects \cite{Cheng14a,Cheng15,Mikhailov16a} and a nonperturbative quasi-classical theory of the nonlinear electrodynamic response of graphene \cite{Mikhailov17a} have been developed.

The optical Kerr effect (OKE) is a nonlinear phenomenon related to a change of the refractive index $n(\omega)=\sqrt{\epsilon(\omega)}$ of a \textit{bulk, three-dimensional} (3D) material in the field of a strong electromagnetic wave, 
\be 
n(\omega)=n_0(\omega)+n_2(\omega)I;\label{n}
\ee 
here $\epsilon(\omega)$ is the dielectric function of the material, and $\omega$ and $I$ are the frequency and the intensity of the wave. The nonlinear refractive index $n_2(\omega)$ is related to the third-order electric susceptibility $\chi^{(3)}(\omega,\omega,-\omega)$ of the three-dimensional (3D) medium,
\ba 
\left[n_0(\omega)+n_2(\omega)I\right]^2\approx n_0^2(\omega)+2n_0(\omega)n_2(\omega)I &&
\nonumber \\ =
1+4\pi\left(\chi^{(1)}(\omega)+\frac 34 \chi^{(3)}(\omega,\omega,-\omega)|E|^2\right);&&\label{eps}
\ea
here $E$ is the electric field of the wave. The functions $\epsilon(\omega)$, $n(\omega)$, $\chi(\omega)$ in Eqs. (\ref{n}) and (\ref{eps}) are, in general, complex. In a weakly absorbing medium the real part of $n_2(\omega)$ is proportional to the real part of $\chi^{(3)}(\omega,\omega,-\omega)$, 
\be 
\textrm{Re }n_2(\omega)=\frac {3\pi}{n_0^2 c} \textrm{Re }\chi^{(3)}(\omega,\omega,-\omega).\label{n2}
\ee
The imaginary part of $n_2(\omega)$ determines the nonlinear absorption and is related to the saturable absorption effect. In the more general case of a non-vanishing absorption the relation between the complex $n_2(\omega)$ and $\chi^{(3)}(\omega,\omega,-\omega)$ is more complicated; it can be found in Ref. \cite{delCoso04}. 
Experimentally the nonlinear refractive index $n_2$ in 3D (bulk) materials (both its real and imaginary parts) can be measured by the $Z$-scan technique \cite{Sheik90}. 

In graphene the optical Kerr and the saturable absorption effects have been experimentally studied in several publications, see, e.g., Refs. \cite{Zhang12,Chen13,Miao15,Dremetsika16,Vermeulen16,Dremetsika17} and \cite{Bao09,Zhang09,Popa10,Popa11,Zheng12}, respectively. Apart from the fundamental interest, these, as well as closely related four-wave mixing phenomena, attract much attention due to a number of their potential photonic and optoelectronic applications such as, for example, the mode locking of lasers \cite{Zhang09,Popa10,Popa11}, frequency conversion\cite{Hendry10,Gu12,KonigOtto17,Alexander17}, and all-optical signal generation and processing \cite{Koos09,Moss13,Chen15}.

To measure the OKE in graphene, different experimental techniques have been used, including $Z$-scan \cite{Zhang12,Chen13,Miao15,Dremetsika16} and the optical heterodyne detection (OHD) scheme \cite{Dremetsika16,Dremetsika17}. Results of these works have been presented in terms of the \textit{effective} nonlinear refractive index $n_2$ of graphene and are rather contradictory. Not only does the absolute value of the measured $n_2$ differ by up to 3 orders of magnitude in different papers, but even about the sign of $n_2$ there still exists no consensus; see a discussion in Ref. \cite{Dremetsika16}.

This situation shows that a detailed analysis of experimental methods of observing OKE in graphene and graphene related materials is highly desirable. Indeed, first of all it should be emphasized that the nonlinear refractive index $n_2$ and other physical quantities [$\epsilon(\omega)$, $\chi(\omega)$] introduced in Eqs. (\ref{n}) and (\ref{eps}) have a clear physical meaning only in bulk, 3D materials. Their definition in the macroscopic electrodynamics \cite{Landau8} implies a procedure of averaging electric fields over ``physically infinitesimal'' volume elements, which means that all sample dimensions should substantially exceed the inter-atomic distance. In graphene and other one- or few-atoms thick \textit{``two-dimensional''} (2D) materials \cite{Novoselov05Nat} the nonlinear refractive index $n_2$ cannot therefore be mathematically rigorously defined, although it is commonly used in experimental papers. Instead, the experimentally measured quantities should be related to the surface (2D) third-order conductivity $\sigma^{(3)}_{\alpha\beta\gamma\delta}(\omega_1,\omega_2,\omega_3)$ which has a clear physical meaning and should therefore be used in the nonlinear graphene (and other 2D crystals) electrodynamics. 

Second, the third conductivity $\sigma^{(3)}_{\alpha\beta\gamma\delta}(\omega_1,\omega_2,\omega_3)$ is a fourth-rank tensor which has several independent nonzero components. It may happen that in different methods different combinations of $\sigma^{(3)}_{\alpha\beta\gamma\delta}$ components are measured. This additionally shows that the OKE in graphene cannot be adequately described by a single scalar quantity $n_2$. 

Third, the measurements in Refs. \cite{Zhang12,Chen13,Miao15,Dremetsika16,Vermeulen16,Dremetsika17} have been performed at a few isolated frequencies (typically at the telecommunication wavelength $\lambda\simeq 1.55$ $\mu$m) and in a nominally undoped graphene, while the theory \cite{Cheng15,Mikhailov16a} predicts a rich behavior of the third conductivity components \textit{as a function of} frequency $\omega$ and Fermi energy $E_F$, with several resonances related to the one-, two-, and three-photon interband transitions. Thus the question arises as to whether and how all the (nonzero) components of $\sigma^{(3)}_{\alpha\beta\gamma\delta}(\omega_1,\omega_2,\omega_3)$ can be extracted from the OKE experiments and which dependencies (on the radiation wavelength, doping, temperature, etc.) are to be expected to be seen in experiments.

In this paper we perform a detailed theoretical analysis of an OHD-OKE experiment in a 2D nonlinear material, derive formulas relating the experimentally measured quantities to the real and imaginary parts of its first- and third-order conductivities and show how all the nonzero OKE-relevant components of the tensor $\sigma^{(3)}_{\alpha\beta\gamma\delta}(\omega_1,\omega_2,\omega_3)$ can be extracted from the OHD-OKE measurements. Within the model of $\sigma^{(3)}_{\alpha\beta\gamma\delta}(\omega_1,\omega_2,\omega_3)$ of graphene derived in Refs. \cite{Cheng15,Mikhailov16a} we analyze its theoretically expected OHD-OKE response in dependence of frequency, Fermi energy, temperature, relaxation rate, and ellipticity of the incident light. 

\section{Analysis of the OHD-OKE experiment: General theory\label{sec:theory}}

For simplicity, we will consider a single graphene layer lying in the plane $z=0$, without any substrate. The influence of different types of substrates on the Kerr response is briefly discussed in Sec. \ref{sec:conclus}.

\subsection{Which $\sigma^{(3)}_{\alpha\beta\gamma\delta}$ components are relevant for OKE?}

In general, the third-order nonlinear response of graphene is determined by the fourth-rank tensor $\sigma^{(3)}_{\alpha\beta\gamma\delta}(\omega_1,\omega_2,\omega_3)$ which has eight (out of sixteen) nonzero complex-valued components depending on three input frequencies $\omega_1$, $\omega_2$, and $\omega_3$. The tensor $\sigma^{(3)}_{\alpha\beta\gamma\delta}(\omega_1,\omega_2,\omega_3)$ satisfies certain symmetry relations \cite{Cheng15,Mikhailov16a}, in particular, simultaneous permutations of the indexes $\beta$, $\gamma$, $\delta$, and the corresponding arguments $\omega_1$, $\omega_2$, $\omega_3$ do not change it, e.g., 
\be  
\sigma^{(3)}_{\alpha\beta\gamma\delta}(\omega_1,\omega_2,\omega_3)=
\sigma^{(3)}_{\alpha\gamma\beta\delta}(\omega_2,\omega_1,\omega_3).\label{symm1}
\ee
Three of the eight nonzero components of the tensor $\sigma^{(3)}_{\alpha\beta\gamma\delta}$ are independent,
\ba 
&&\sigma^{(3)}_{xxyy}(\omega_1,\omega_2,\omega_3)= \sigma^{(3)}_{yyxx}(\omega_1,\omega_2,\omega_3), \nonumber \\ &&\sigma^{(3)}_{xyxy}(\omega_1,\omega_2,\omega_3)= \sigma^{(3)}_{yxyx}(\omega_1,\omega_2,\omega_3), \\ &&\sigma^{(3)}_{xyyx}(\omega_1,\omega_2,\omega_3)= \sigma^{(3)}_{yxxy}(\omega_1,\omega_2,\omega_3),\nonumber 
\label{indep_components}
\ea 
and the component $\sigma^{(3)}_{xxxx}(\omega_1,\omega_2,\omega_3)= \sigma^{(3)}_{yyyy}(\omega_1,\omega_2,\omega_3)$ is the sum of the other three, \ba 
\sigma^{(3)}_{xxxx}(\omega_1,\omega_2,\omega_3)&=& \sigma^{(3)}_{xxyy}(\omega_1,\omega_2,\omega_3)+ \sigma^{(3)}_{xyxy}(\omega_1,\omega_2,\omega_3)\nonumber \\ &+& \sigma^{(3)}_{xyyx}(\omega_1,\omega_2,\omega_3).\label{indep_components-2}
\ea

The relations (\ref{symm1}) -- (\ref{indep_components-2}) are valid for all third-order nonlinear effects. The OKE is a special case determined by the functions $\sigma^{(3)}_{\alpha\beta\gamma\delta}(\omega,\omega,-\omega)$. In this case only two nonzero components are independent, since according to (\ref{symm1}), 
\be 
\sigma^{(3)}_{xxyy}(\omega,\omega,-\omega)=
\sigma^{(3)}_{xyxy}(\omega,\omega,-\omega).\label{symm2}
\ee
We will express all our results via two independent components $\sigma^{(3)}_{xxxx}(\omega,\omega,-\omega)$ and $\sigma^{(3)}_{xxyy}(\omega,\omega,-\omega)$ of the $\sigma^{(3)}$ tensor. The third nonzero component of the $\sigma^{(3)}$ tensor can then be found from the relation  
\be 
\sigma^{(3)}_{xyyx}(\omega,\omega,-\omega) =\sigma^{(3)}_{xxxx}(\omega,\omega,-\omega)-
2\sigma^{(3)}_{xxyy}(\omega,\omega,-\omega).
\label{sigmaSrelation}
\ee
Below we aim to find the relations between the experimentally measured quantities and the real and imaginary parts of the complex functions $\sigma^{(3)}_{xxxx}(\omega,\omega,-\omega)$ and $\sigma^{(3)}_{xxyy}(\omega,\omega,-\omega)$. 

\subsection{Derivation of the main formulas
\label{sec:gen-theory}}

\begin{figure*}
\includegraphics[width=0.98\columnwidth]{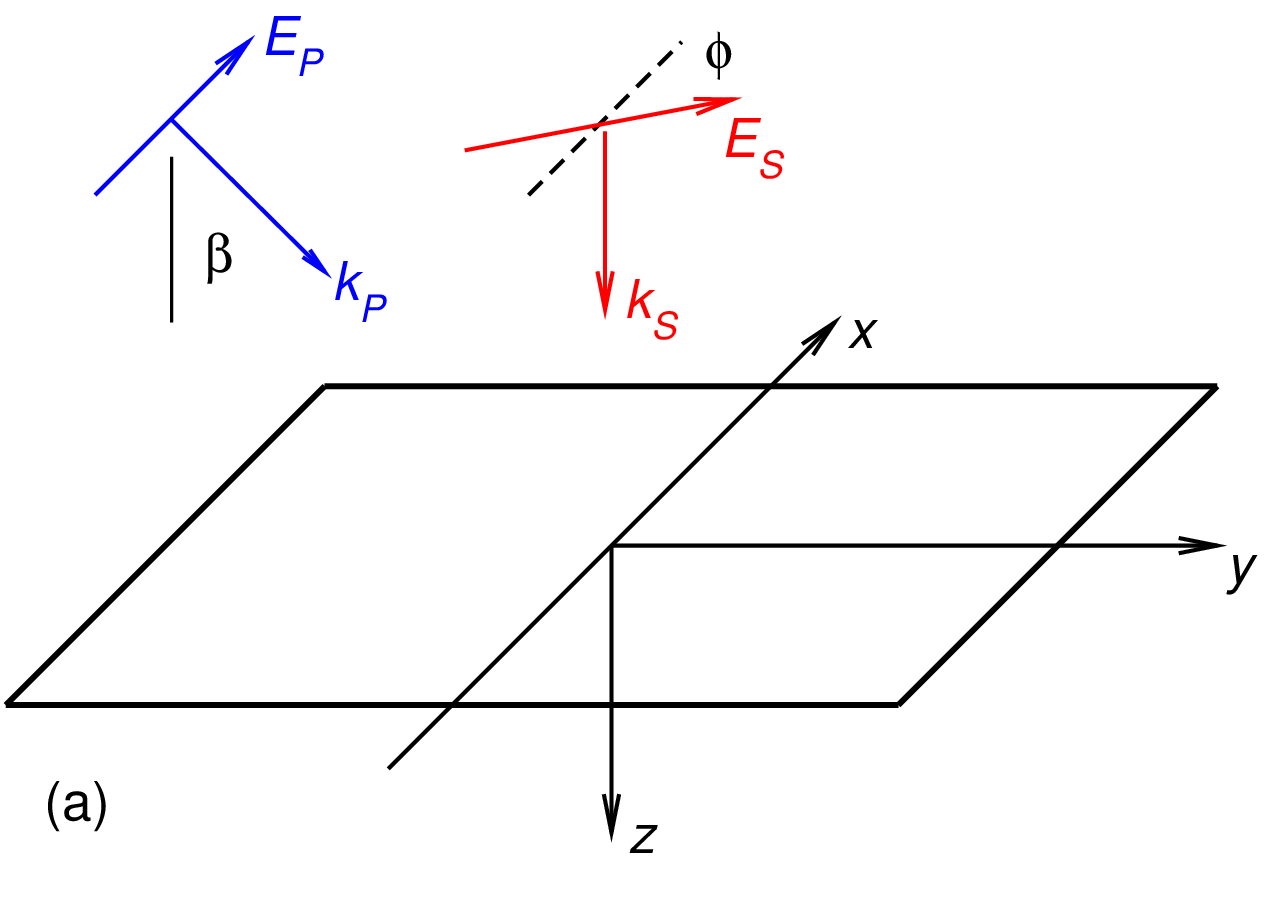}
\includegraphics[width=0.98\columnwidth]{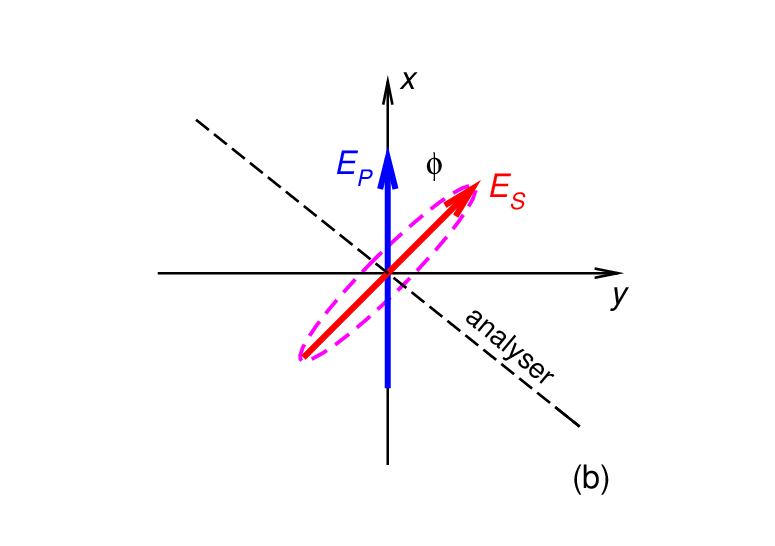}
\caption{\label{fig:geom}The geometry of the OHD measurements setup used in Ref. \cite{Dremetsika16}: (a) The pump wave $P$ is incident upon the graphene layer lying at the plane $z=0$ and is linearly polarized in the $x$-direction. The incidence angle is $\beta$. (b) The polarization plane of the probe signal $S$ is rotated by the angle $\phi$ relative to the $x$ axis. The probe wave $S$ can be elliptically polarized with the ellipticity determined by the parameter $\theta$. The dashed line in (b) shows the transmission axis of the analyzer placed between the graphene layer and the detector.}
\end{figure*}

In a typical OHD-OKE experiment, see, e.g., Ref. \cite{Dremetsika16}, two different waves, the pump wave ($P$) and the probe wave (which we will call ``signal,'' $S$, to designate the two waves by short different subscripts) are incident on graphene lying in the plane $z=0$, Fig. \ref{fig:geom}(a). The pump ($P$) wave is incident on the graphene plane under the angle $\beta$ and is linearly polarized in the $x$ direction, 
\ba 
{\bm E}^{\rm ext}_P(y,z,t)&=&\frac {E_Pe^{i(\omega_P/c)y\sin\beta }}2
\left(
\begin{array}{c}
1\\ 0\\
\end{array}
\right)\nonumber \\ &\times& e^{i(\omega_P/c)z\cos\beta -i\omega_P t}+ \textrm{c.c.}\label{Ep}
\ea
The probe ($S$) wave is normally incident on the graphene plane and is linearly polarized under the angle $\phi$ to the polarization of the $P$ wave. The $S$ wave can also be elliptically polarized with the ratio of the short to long axes determined by $\tan\theta$, see Fig. \ref{fig:geom}(b). The field of the incident probe wave can thus be written as
\ba 
{\bm E}^{\rm ext}_S(z,t)=\frac{E_Se^{i\psi}}{2}\left(
\begin{array}{c}
\cos\phi e^{i\theta}\\ \sin\phi e^{-i\theta}\\
\end{array}
\right)
e^{i\omega_Sz/c-i\omega_S t}+ \textrm{c.c.} ,\nonumber \\\label{Es}
\ea
where the amplitudes $E_P$ and $E_S$ are assumed to be real and the phase angle $\psi$ takes into account a possible phase shift between the pump and probe waves. 
If $\theta>0$ the form (\ref{Es}) corresponds to the $\bm E_S$ vector rotating in the counterclockwise direction if to look in the wave propagation (positive $z$-) direction. The frequencies of the pump and probe waves were the same in the OKE experiment, $\omega_P=\omega_S=\omega$.

The fields ${\bm E}_P^{\rm ext}$ and ${\bm E}_S^{\rm ext}$ in Eqs. (\ref{Ep}) and (\ref{Es}) are the fields of the \textit{external} waves incident on the graphene layer. The fields actually acting on graphene electrons at $z=0$ differ from the external fields and can be found by solving Maxwell equations in the linear order. The result for the fields at $z=0$ is
\be 
{\bm E}_P(y,0,t)=\frac{E_Pe^{i(\omega/c) y\sin\beta }}{2\left(1+\frac{2\pi\sigma^{(1)}_{\omega}}{c\cos\beta}\right)}
\left(
\begin{array}{c}
1\\ 0\\
\end{array}
\right)e^{-i\omega t}+ \textrm{c.c.},\label{EpActing}
\ee
\be 
{\bm E}_S(0,t)=\frac{E_Se^{i\psi}} {2\left(1+\frac{2\pi\sigma^{(1)}_{\omega}}{c}\right)}
\left(
\begin{array}{c}
\cos\phi e^{i\theta}\\ \sin\phi e^{-i\theta}\\
\end{array}
\right)e^{-i\omega t}+ \textrm{c.c.}, \label{EsActing}
\ee
where $\sigma^{(1)}_{\omega}$ is the linear (first-order) conductivity of graphene \cite{Mikhailov07d} and the factors $\left(1+2\pi\sigma^{(1)}_{\omega}/c\right)$ in the denominators are due to the self-consistent screening effect. The linear conductivity $\sigma^{(1)}_{\omega}$ is in general a complex function; its real part is responsible for the linear absorption in graphene.

The fields (\ref{EpActing})--(\ref{EsActing}) should then be substituted in the third-order current $j_{\alpha}^{(3)}(t)$, 
\ba
j_{\alpha}^{(3)}(t)&=&
\int_{-\infty}^\infty d\omega_1 
\int_{-\infty}^\infty d\omega_2  
\int_{-\infty}^\infty d\omega_3 
\sigma^{(3)}_{\alpha\beta\gamma\delta}(\omega_1,\omega_2,\omega_3)
\nonumber \\ &\times&
E^\beta_{\omega_1} 
E^\gamma_{\omega_2} 
E^\delta_{\omega_3} 
e^{-i(\omega_1 +\omega_2+\omega_3) t},
\label{sigma3}
\ea
where $E^\beta_{\omega_1}$, $E^\gamma_{\omega_2}$, $E^\delta_{\omega_3}$ are Fourier components of the fields (\ref{EpActing})--(\ref{EsActing}) and $\sigma^{(3)}_{\alpha\beta\gamma\delta}(\omega_1,\omega_2,\omega_3)$ is the fourth-rank conductivity tensor of graphene calculated in Refs. \cite{Cheng15,Mikhailov16a}, see, e.g., Eqs. (59)--(78) in \cite{Mikhailov16a}. Notice that by ignoring the wave-vector arguments $\bm q_1$, $\bm q_2$, etc., in the function $\sigma^{(3)}_{\alpha\beta\gamma\delta}(\omega_1,\omega_2,\omega_3)$ we assume that the external field is approximately uniform in the plane of the 2D layer, and the nonlocal effects in the third order can be ignored. This implies that the angle $\beta$ in Eqs. (\ref{Ep}) and (\ref{EpActing}) should be sufficiently small. The required smallness of $\beta$ is quantitatively determined by the condition $(\omega/ck_F)\sin\beta\ll 1$, where $k_F$ is the Fermi wave-vector. This condition is usually satisfied in the experiments.

Substituting the Fourier components of the fields (\ref{EpActing}) and (\ref{EsActing}) into the third-order current (\ref{sigma3}) we get a sum of a large number of terms. Taking into account only those that lead to the wave propagating in the $z$ direction toward the detector (i.e. only the $y$-independent contributions) we get
\ba 
&&{\bm j}_{\rm unif}^{(3)}(t)=
3\Bigg\{
2\left(
\begin{array}{c}
\sigma^{(3)}_{xxxx}(\omega,\omega,-\omega) E_{xS} \\
\sigma^{(3)}_{yyxx}(\omega,\omega,-\omega)E_{yS}\\
\end{array}
\right)|E_{xP}|^2 
\nonumber \\ &+&
\left(
\begin{array}{c}
\sigma^{(3)}_{xxxx}(\omega,\omega,-\omega)  E_{xS}\\
2\sigma^{(3)}_{yyxx}(\omega,\omega,-\omega)  
 E_{yS}\\
\end{array} 
\right)|E_{xS}|^2
\nonumber \\ &+&
\left(
\begin{array}{c}
\sigma^{(3)}_{xyyx}(\omega,\omega,-\omega) E_{xS}^\star E_{yS}^2 \\
\sigma^{(3)}_{yxxy}(\omega,\omega,-\omega) E_{xS}^2 E_{yS}^\star \\
\end{array}
\right)
\nonumber \\ &+&
\left(
\begin{array}{c}
2\sigma^{(3)}_{xxyy}(\omega,\omega,-\omega) E_{xS} \\
\sigma^{(3)}_{yyyy}(\omega,\omega,-\omega) E_{yS} \\
\end{array}
\right)|E_{yS}|^2
\Bigg\}e^{-i\omega t},\label{unif-curr}
\ea
where we have omitted the complex conjugate terms, $E_{xP}$, $E_{xS}$, and $E_{yS}$ are the complex field components from Eqs. (\ref{EpActing}) and (\ref{EsActing}) and the subscript ``unif'' reminds us that only the uniform ($y$-independent) contributions to the third-order current are included in (\ref{unif-curr}). By calculating the electric field of the wave emitted by the oscillating third-order current (\ref{unif-curr}), 
\be 
{\bm E}_{\rm unif}^{(3)}(t)= - \frac{2\pi/c} {1+2\pi\sigma^{(1)}_{\omega}/c}{\bm j}_{\rm unif}^{(3)}(t),
\ee
and adding it to the field (\ref{EsActing}) of the linear wave passing through the graphene layer we obtain the total electric field of the wave (including the first and third order) passing through the graphene layer and propagating towards the detector:
\begin{widetext}
\ba 
{\bm E}_{\rm unif}^{(1)+(3)}(z,t)&=&
\frac{E_Se^{i\omega z/c-i\omega t+i\psi}}{2\left(1+\frac{2\pi\sigma^{(1)}_{\omega}}{c}\right)}
\Bigg\{\left(
\begin{array}{c}
\cos\phi e^{i\theta} \\
\sin\phi e^{-i\theta}\\
\end{array}
\right)
-
\frac{ \frac {3\pi}{2c}}{1+\frac{2\pi\sigma^{(1)}_{\omega}}{c}}
\Bigg[
2\left(
\begin{array}{c}
\sigma^{(3)}_{xxxx}(\omega,\omega,-\omega) \cos\phi e^{i\theta} \\
\sigma^{(3)}_{yyxx}(\omega,\omega,-\omega)\sin\phi e^{-i\theta}\\
\end{array}
\right)
\frac{|E_P|^2}{\left|1+\frac{2\pi\sigma^{(1)}_{\omega}} {c\cos\beta}\right|^2}
\nonumber \\ &+&
\left(
\begin{array}{c}
\cos\phi e^{i\theta}
\left(\sigma^{(3)}_{xxxx}(\omega,\omega,-\omega)  +\sin^2\phi 
\sigma^{(3)}_{xyyx}(\omega,\omega,-\omega) 
 \left(e^{-i4\theta}-1\right) \right)
\\
\sin\phi e^{-i\theta}
\left(\sigma^{(3)}_{xxxx}(\omega,\omega,-\omega)+
\cos^2\phi
 \sigma^{(3)}_{xyyx}(\omega,\omega,-\omega) 
 \left(e^{i4\theta} 
-1\right)
  \right) \\
\end{array} 
\right) 
\frac{|E_S|^2}{\left|1+\frac{2\pi\sigma^{(1)}_{\omega}}{c}\right|^2}\Bigg]
\Bigg\}.
\label{field}
\ea
Calculating now the projection of the field (\ref{field}) on the transmission axis of the analyzer we get the field of the wave registered by the detector
\ba 
E_{{\rm detect}}(t)&=&
\frac{E_S\sin\phi \cos\phi }{\left(1+\frac{2\pi\sigma^{(1)}_{\omega}}{c}\right)} e^{i\omega z/c-i\omega t+i\psi} \Bigg\{-i\sin\theta\Bigg|_{\textrm{term }A} 
+
|E_P|^2\frac {3\pi}{2c}\frac{\Big(
\sigma^{(3)}_{xxxx}(\omega,\omega,-\omega)  e^{i\theta} -
\sigma^{(3)}_{xxyy}(\omega,\omega,-\omega) e^{-i\theta}
\Big)}{\left(1+\frac{2\pi\sigma^{(1)}_{\omega}}{c}\right) \left|1+\frac{2\pi\sigma^{(1)}_{\omega}} {c\cos\beta}\right|^2}\Bigg|_{\textrm{term }B}
\nonumber \\ &+&
\frac{\frac {3\pi}{4c}|E_S|^2} {\left(1+\frac{2\pi\sigma^{(1)}_{\omega}}{c}\right)\left|1+\frac{2\pi\sigma^{(1)}_{\omega}}{c}\right|^2}
\Biggl[
\Big(
\sigma^{(3)}_{xxxx}(\omega,\omega,-\omega)  e^{i\theta}-
 2\sigma^{(3)}_{xxyy}(\omega,\omega,-\omega)  e^{-i\theta}
-
\sigma^{(3)}_{xyyx}(\omega,\omega,-\omega) 
 e^{i3\theta}\Big)\cos^2\phi 
\nonumber \\ &-&
\Big(
\sigma^{(3)}_{xxxx}(\omega,\omega,-\omega)   e^{-i\theta} 
-2\sigma^{(3)}_{xxyy}(\omega,\omega,-\omega) 
  e^{i\theta} -
\sigma^{(3)}_{xyyx}(\omega,\omega,-\omega) 
 e^{-i3\theta}  \Big)\sin^2\phi
\Biggr]\Bigg|_{\textrm{term }C}\Bigg\}.\label{field-detect}
\ea
\end{widetext}
It contains three terms, the linear one proportional to $E_S$ (the term $A$) and two nonlinear terms proportional to $E_S|E_P|^2$ and $E_S|E_S|^2$ (the terms $B$ and $C$, respectively). The intensity of the wave entering the detector then contains six contributions, 
\be 
I_{\rm detect}= \frac c{8\pi}|E_{{\rm detect}}(t)|^2 =\sum_{J={\rm I}}^{\rm VI}I_{\rm detect}^J
\ee
which we write down assuming that $\phi=\pi/4$ (see Ref. \cite{Dremetsika16}). The first term,
\be 
I_{\rm detect}^{\rm I}=I_S \frac{\sin^2\theta }{4\left|1+\frac{2\pi\sigma^{(1)}_{\omega}}{c}\right|^2}  ,
\label{intensI}
\ee
results from the squared term $A$ in (\ref{field-detect}) and is the linear one. It is just the probe ($S$) wave which reaches the detector if the ellipticity of the wave $\theta$ is not zero. Here and below the quantities 
\be 
I_P=\frac c{8\pi}|E_P|^2, \ \ \ \ I_S=\frac c{8\pi}|E_S|^2
\ee
are the intensities of the incident pump and probe (signal) waves. 

All other terms contain components of the third-order conductivity tensor. The second one,
\be 
I_{\rm detect}^{\rm II}= \frac {6\pi^2}{c^2} \frac{I_SI_P\sin\theta}{\left|1+\frac{2\pi\sigma^{(1)}_{\omega}}{c}\right|^2 \left|1+\frac{2\pi\sigma^{(1)}_{\omega}} {c\cos\beta}\right|^2}    
{\rm Im } \left(
\frac{{\cal U}(\omega,\theta)
}{1+\frac{2\pi\sigma^{(1)}_{\omega}}{c} }
\right),
\label{intensII}
\ee
is due to the interference of the $A$ and $B$ terms in Eq. (\ref{field-detect}) and contains a certain linear combination of $\sigma^{(3)}_{xxxx}$ and $\sigma^{(3)}_{xxyy}$ components, see Eq. (\ref{funcU}). It is of the second order (proportional to $I_SI_P$) and is finite if the ellipticity $\theta$ is not zero. 
The third term,
\ba 
I_{\rm detect}^{\rm III}=\frac {6\pi^2}{c^2}
 \frac{I_S^2\sin^2\theta }{\left|1+\frac{2\pi\sigma^{(1)}_{\omega}}{c}\right|^4}   
 {\rm Re }\left(
\frac{ {\cal V}(\omega,\theta)  } {1+\frac{2\pi\sigma^{(1)}_{\omega}}{c}}
\right),
\label{intensIII}
\ea
is due to the interference of the $A$ and $C$ terms in Eq. (\ref{field-detect}). It is also of the second order (proportional to $I_S^2$) and disappears at $\theta=0$. It contains a different linear combination of the same components of the $\sigma^{(3)}$ tensor, see Eq. (\ref{funcV}).

The remaining three terms are of the third order. The fourth and fifth ones are determined by the squared terms $B$ and $C$ in Eq. (\ref{field-detect}): 
\ba 
I_{\rm detect}^{\rm IV}&=& \frac {36\pi^4}{c^4}I_SI_P^2
 \frac{\left|{\cal U}(\omega,\theta)
\right|^2}{\left|1+\frac{2\pi\sigma^{(1)}_{\omega}}{c}\right|^4 \left|1+\frac{2\pi\sigma^{(1)}_{\omega}} {c\cos\beta}\right|^4},
\label{intensIV}
\ea

\ba 
I_{\rm detect}^{\rm V}&=&\frac {36\pi^4}{c^4}I_S^3 \sin^2\theta 
\frac{\left| 
{\cal V}(\omega,\theta)
\right|^2}{\left|1+\frac{2\pi\sigma^{(1)}_{\omega}}{c}\right|^8};
\label{intensV}
\ea
the term IV remains finite in the limit $\theta=0$ (at the linear polarization of the probe wave). 
The last, sixth term is due to the interference of the $B$ and $C$ terms in Eq. (\ref{field-detect}),
\ba 
I_{\rm detect}^{\rm VI}&=&\frac {72\pi^4}{c^4}
\frac{I_S^2I_P\sin\theta}
{\left|1+\frac{2\pi\sigma^{(1)}_{\omega}}{c}\right|^6
\left|1+\frac{2\pi\sigma^{(1)}_{\omega}} {c\cos\beta}\right|^2} 
\nonumber \\ &\times& 
{\rm Im } \Big( 
{\cal U}(\omega,\theta)
{\cal V}^\star(\omega,\theta) \Big);
\label{intensVI}
\ea
the star means the complex conjugate. 

The analytical formulas (\ref{intensII}) -- (\ref{intensVI}) representing five nonlinear contributions II -- VI is the main result of this work valid for any 2D material. Apart from the parameter $2\pi\sigma_\omega^{(1)}/c$ which can be determined from the linear graphene response, two combinations of the components of $\sigma^{(3)}_{\alpha\beta\gamma\delta}$,
\be 
{\cal U}(\omega,\theta)=-\left[
\sigma^{(3)}_{xxxx}(\omega,\omega,-\omega)  e^{i\theta} -
\sigma^{(3)}_{xxyy}(\omega,\omega,-\omega) e^{-i\theta}\right]
\label{funcU}
\ee
and 
\ba
{\cal V}(\omega,\theta)&=& \sigma^{(3)}_{xxxx}(\omega,\omega,-\omega) \cos 2\theta 
 \nonumber \\ & -&
 2\sigma^{(3)}_{xxyy}(\omega,\omega,-\omega)\left( 1+\cos 2\theta 
 \right) 
\label{funcV}
\ea 
enter these formulas. Independently measuring the contributions II -- VI one can determine the real and imaginary parts of ${\cal U}$ and ${\cal V}$, and hence, the real and imaginary parts of $\sigma^{(3)}_{xxxx}(\omega,\omega,-\omega)$ and $\sigma^{(3)}_{xxyy}(\omega,\omega,-\omega)$. The components $\sigma^{(3)}_{xyyx}(\omega,\omega,-\omega)$ and $\sigma^{(3)}_{xyxy}(\omega,\omega,-\omega)$ can then be found using Eqs. (\ref{sigmaSrelation}) and (\ref{symm2}). 

\subsection{Measurements at different modulation frequencies}

In order to independently measure contributions II -- VI to the detected signal, one can use an elegant method employed in Ref. \cite{Dremetsika16}. In that experiment the intensities of the incident pump and probe waves were modulated with different frequencies, $f_P$ and $f_S$ (in Ref. \cite{Dremetsika16} $f_P=5f_w=205$ Hz and $f_S=6f_w=246$ Hz with $f_w=41$ Hz). Assuming for simplicity that the modulation was (or can be made) sinusoidal, we write the pump and probe signal amplitudes as
\be 
I_P=I_P^0[1+\alpha_P\cos(2\pi f_Pt)],\ I_S=I_S^0[1+\alpha_S\cos(2\pi f_St)],
\label{modfreq}
\ee
where $\alpha_{P,S}$ are the modulation depths. The output signal registered by the detector then contains a number of different harmonics listed in Table \ref{tab:tableHarm}. One sees that all third-order terms, IV, V, and VI, can be uniquely measured at one of the output modulation frequencies, i.e., term IV at one of the frequencies $2f_P$, $f_S+ 2f_P$, or $f_S- 2f_P$, term V at frequency $3f_S$, and term VI at one of the frequencies $2f_S+ f_P$ or $2f_S- f_P$. 

\begin{table*}[t]
\caption{\label{tab:tableHarm}
Modulation frequencies for different intensity contributions to the detected signal. The last column shows numerical values of the frequencies for the data from Ref. \cite{Dremetsika16} ($f_P=5f_w=205$ Hz and $f_S=6f_w=246$ Hz with $f_w=41$ Hz). Underlined are the frequencies which occur only once in the table.}
\begin{ruledtabular}
\begin{tabular}{cccc}
\textrm{Contribution}& Proportional to &
\textrm{Modulation frequency} &  in Ref. \cite{Dremetsika16} (in Hz)\\
\colrule
I & $I_S$ &  $f_S$ & 246 \\
II & $I_SI_P$  & $f_S$, $f_P$, $f_S\pm f_P$ & 246, 205, 451, 41 \\
III & $I_S^2$  & $f_S$, $2f_S$ & 246, 492 \\
IV & $I_SI_P^2$  & $f_S$, $f_P$, $f_S\pm f_P$, $\underline{2f_P}$, $\underline{ f_S\pm 2f_P}$ & 246, 205, 451, 41, \underline{410, 656, 164} \\
V & $I_S^3$ & $f_S$, $2f_S$, $\underline{3f_S}$ & 246, 492, \underline{738}\\
VI & $I_S^2I_P$  & $f_S$, $f_P$, $2f_S$, $f_S\pm f_P$, $\underline{2f_S\pm f_P}$ & 246, 205, 492, 451, 41, \underline{697, 287}\\
\end{tabular}
\end{ruledtabular}
\end{table*}

For other contributions the measurements are not so unambiguous. For example, by measuring the output signal at the sum or difference frequency $f_S\pm f_P$ one gets the contributions from terms II, IV, and VI. The signal measured at the modulation frequency $2f_S$ contains contributions from terms III, V, and VI. In these cases one should use additional dependencies, e.g., on the ellipticity $\theta$ or on the intensities $I_P$ or $I_S$, in order to unambiguously extract  terms II--VI from the output signal. 

\subsection{What was measured in Ref. \cite{Dremetsika16}?\label{sec:whatwas}}

A detailed study of the OKE using the OHD technique was performed in Ref. \cite{Dremetsika16}. In that paper the output signal was detected at the sum of the modulation frequencies $f_S+ f_P$, and the authors presented the experimentally measured wave intensity as a sum of two contributions, 
\ba 
I_{\rm det, exp}^{f_S+ f_P}(\theta)&=&\left|\frac{2\pi }{\lambda} \left(n_2^{\rm eff}+i\kappa_2^{\rm eff}\right)L^{\rm eff}\right|^2 I_P^2I_S
\nonumber \\ &+& \sin \theta \frac{2\pi }{\lambda} n_2^{\rm eff}L^{\rm eff} I_PI_S,\label{exper}
\ea
where the first (designated as ``homo'' in \cite{Dremetsika16}) term is proportional to $I_P^2I_S$ and was assumed to be $\theta$ independent and the second (designated as ``hetero'') term is proportional to $\theta I_PI_S$ at $\theta\ll 1$; here $\lambda=1600$ nm is the radiation wavelength, $L^{\rm eff}\simeq 0.33$ nm is the \textit{effective} graphene thickness, and we have supplied quantities which are poorly defined in 2D systems (see Sec. \ref{sec:Intro}) by the superscript ``eff''. Now, combining all our terms (II, IV, and VI) which contribute to the output signal  at the modulation frequency $f_S+ f_P$ we obtain
\begin{widetext}
\ba 
I_{\rm det, th}^{f_S+ f_P}(\omega,\theta)&=&I_S^0I_P^0\alpha_P\alpha_S
\left(\frac 12\frac {6\pi^2}{c^2} \frac{\sin\theta}{\left|1+\frac{2\pi\sigma^{(1)}_{\omega}}{c}\right|^2 \left|1+\frac{2\pi\sigma^{(1)}_{\omega}} {c\cos\beta}\right|^2} 
\left(\frac{{\cal U}(\omega,\theta)}{1+\frac{2\pi\sigma^{(1)}_{\omega}}{c}}\right)''
\right.
\nonumber \\ &+&
\left.
\frac {36\pi^4}{c^4}I_P^0
\frac{\left|{\cal U}(\omega,\theta)\right|^2}{\left|1+\frac{2\pi\sigma^{(1)}_{\omega}}{c}\right|^4 \left|1+\frac{2\pi\sigma^{(1)}_{\omega}} {c\cos\beta}\right|^4}+
\frac {72\pi^4}{c^4}I_S^0\frac{\sin\theta} {\left|1+\frac{2\pi\sigma^{(1)}_{\omega}}{c}\right|^6 \left|1+\frac{2\pi\sigma^{(1)}_{\omega}} {c\cos\beta}\right|^2}
\Big( 
{\cal U}(\omega,\theta)
{\cal V}^\star(\omega,\theta) \Big)''\right).\label{SignalAtSum}
\ea
The first term in brackets here evidently corresponds to the ``hetero'' contribution in (\ref{exper}), the second term to the ``homo'' contribution, and the third one was ignored in Ref. \cite{Dremetsika16} (in the experiment $I_S\ll I_P$). Notice also that in the theory the ``homo'' term is $\theta$ dependent and the $\theta$ dependence of the ``hetero'' term is more complicated than just $\simeq \sin\theta$, due to the function ${\cal U}(\omega,\theta)$; see further discussion of this issue in Sec. \ref{sec:ellip-polariz}.

Further, in Ref. \cite{Dremetsika16} the authors calculated the difference of the measured intensity (\ref{exper}) at $+\theta$ and $-\theta$ and got the quantity proportional to $n_2^{\rm eff}$:
\be 
\Delta_\theta I_{\rm det,exp}^{f_S+ f_P}= \sin \theta \frac{4\pi }{\lambda} n_2^{\rm eff}L^{\rm eff} I_PI_S,\label{experDiff}
\ee
Taking the same difference of the theoretically found intensity (\ref{SignalAtSum}) we obtain
\be
\Delta_\theta I_{\rm det,th}^{f_S+ f_P}(\omega,\theta)=
\left[\Delta_\theta I_{\rm det,th}^{f_S+ f_P}(\omega,\theta)\right]_{\rm II}+\left[\Delta_\theta I_{\rm det,th}^{f_S+ f_P}(\omega,\theta)\right]_{\rm IV+VI},\label{DiffTheta}
\ee
where
\be 
\left[\Delta_\theta I_{\rm det,th}^{f_S+ f_P}(\omega,\theta)\right]_{\rm II} =
-\frac {6\pi^2}{c^2} 
\frac{\alpha_P\alpha_SI_P^0I_S^0\sin\theta}
{\left|1+\frac{2\pi\sigma^{(1)}_{\omega}}{c}\right|^2 \left|1+\frac{2\pi\sigma^{(1)}_{\omega}} {c\cos\beta}\right|^2} 
\left(
\frac{
\sigma^{(3)}_{xxxx}(\omega,\omega,-\omega) -
\sigma^{(3)}_{xxyy}(\omega,\omega,-\omega) }
{1+\frac{2\pi\sigma^{(1)}_{\omega}}{c}}\right)'' ,
\label{DiffThetaII}
\ee
and
\ba 
&&\left[\Delta_\theta I_{\rm det,th}^{f_S+ f_P}(\omega,\theta)\right]_{\rm IV+VI} =-
\left(\frac {12\pi^2}{c^2} \right)^2
\frac{\alpha_P\alpha_SI_P^0I_S^0\sin\theta}
{\left|1+\frac{2\pi\sigma^{(1)}_{\omega}}{c}\right|^4 \left|1+\frac{2\pi\sigma^{(1)}_{\omega}} {c\cos\beta}\right|^2} 
\left(\frac{2I_P^0}
{\left|1+\frac{2\pi\sigma^{(1)}_{\omega}} {c\cos\beta}\right|^2}
+\frac{3I_S^0} {\left|1+\frac{2\pi\sigma^{(1)}_{\omega}}{c}\right|^2}\right)
\nonumber \\ && \hspace{2cm}\times
\left[\Big(\sigma^{(3)}_{xxxx}(\omega,\omega,-\omega)\Big)' 
\Big(\sigma^{(3)}_{xxyy}(\omega,\omega,-\omega)\Big)''   -
\Big(\sigma^{(3)}_{xxxx}(\omega,\omega,-\omega)   
\Big)''\Big(
\sigma^{(3)}_{xxyy}(\omega,\omega,-\omega)\Big)' \right].
\label{DiffThetaIV+VI}
\ea
The quantity measured in Ref. \cite{Dremetsika16} thus consists of the term (\ref{DiffThetaII}) resulting from  contribution II and two terms (\ref{DiffThetaIV+VI}) resulting from the contributions IV and VI. All three terms are proportional to $\sin\theta\approx\theta$ at small values of the ellipticity $\theta$. As will be seen below (Sec. \ref{sec:comparison}), in different frequency ranges and at different wave intensities the contributions (\ref{DiffThetaIV+VI}) can be both smaller than (\ref{DiffThetaII}), as was assumed in Ref. \cite{Dremetsika16}, and comparable with or even larger than (\ref{DiffThetaII}). Therefore in general the full result (\ref{DiffTheta}) should be used when the theory is compared with experiment.

If we assume now that the terms IV and VI [Eq. (\ref{DiffThetaIV+VI})] are small as compared to term II [Eq. (\ref{DiffThetaII}] (exact conditions for this will be established below), then the right hand sides of Eqs. (\ref{experDiff}) and (\ref{DiffThetaII}) should correspond to each other. Then we get the relation between \textit{effective} quantities $n_2^{\rm eff}$ and $L^{\rm eff}$ and the components of the third-order conductivity tensor:
\be 
n_2^{\rm eff} L^{\rm eff} \ \Leftrightarrow \ 
-\frac {6\pi^2}{\omega c} \frac 1{\left|1+\frac{2\pi\sigma^{(1)}_{\omega}}{c}\right|^2 \left|1+\frac{2\pi\sigma^{(1)}_{\omega}} {c\cos\beta}\right|^2}
\left(
\frac{\sigma^{(3)}_{xxxx}(\omega,\omega,-\omega) -
\sigma^{(3)}_{xxyy}(\omega,\omega,-\omega)}{1+\frac{2\pi\sigma^{(1)}_{\omega}}{c}} \right)'' ;
\label{correspondence2}
\ee
we have put here $\alpha_P=\alpha_S\simeq 1$. 
The real part of the same quantity determines the \textit{effective} nonlinear absorption coefficient $\kappa_2^{\rm eff}$:
\be 
\kappa_2^{\rm eff} L^{\rm eff} \ \Leftrightarrow \ 
\frac {6\pi^2}{\omega c} \frac 1{\left|1+\frac{2\pi\sigma^{(1)}_{\omega}}{c}\right|^2 \left|1+\frac{2\pi\sigma^{(1)}_{\omega}} {c\cos\beta}\right|^2}
\left(
\frac{\sigma^{(3)}_{xxxx}(\omega,\omega,-\omega) -
\sigma^{(3)}_{xxyy}(\omega,\omega,-\omega)} {1+\frac{2\pi\sigma^{(1)}_{\omega}}{c}} \right)' .
\label{correspondence3}
\ee
\end{widetext}
The squared modulus $\left|n_2^{\rm eff}+i\kappa_2^{\rm eff}\right|^2$ can be obtained by measuring the intensity of the ``homo'' term at the linear polarization of the incident probe wave; compare Eqs. (\ref{exper}) and (\ref{SignalAtSum}) at $\theta=0$. 

The factors $|2\pi\sigma^{(1)}_{\omega}/c|\ll 1$ in the denominators of formulas (\ref{correspondence2})--(\ref{correspondence3}) are often small as compared to unity, e.g., at high (IR, optical) frequencies. One sees that, if to neglect them, the complex \textit{effective} nonlinear refractive index is determined by the difference $\sigma^{(3)}_{xxxx}(\omega,\omega,-\omega) -\sigma^{(3)}_{xxyy}(\omega,\omega,-\omega)$. It is important to understand that this statement is valid only for the OHD-OKE experiment. The quantities $n_2^{\rm eff}$ and $\kappa_2^{\rm eff}$ extracted from a different, e.g., $Z$-scan experiment, will be proportional to a different combination of the $\sigma^{(3)}_{\alpha\beta\gamma\delta}$ components [in the simplest case to $\sigma^{(3)}_{xxxx}(\omega,\omega,-\omega)$]; therefore a direct comparison of results of the $Z$-scan and OHD-OKE measurements is inapplicable. 

\section{Analysis of results using model expressions for the third conductivity \label{sec:theory-model}}

The relations derived in Sec. \ref{sec:theory} are general and do not use any specific model of the third conductivity tensor. Now we analyze some of the key formulas obtained above as a function of frequency, doping, temperature, etc.  using the model of $\sigma^{(3)}_{\alpha\beta\gamma\delta}(\omega,\omega,-\omega)$ developed in Refs. \cite{Cheng15,Mikhailov16a}. For the linear and third-order conductivities at temperature $T=0$ we use formulas of Ref. \cite{Mikhailov16a} [Eqs. (44) -- (48) and (59) -- (78) respectively]. For the finite temperature conductivities $\sigma_\omega^{(1)}(\mu,T)$ and $\sigma_{\alpha\beta\gamma\delta}^{(3)}(\omega_1,\omega_2,\omega_3;\mu,T)$ we use the relation \cite{Maldague78}
\be 
\sigma_\omega^{(1)}(\mu,T)=
\frac 1{4T}\int_{-\infty}^\infty \frac{\sigma_\omega^{(1)}(E_F,0)} {\cosh^2\left(\frac{\mu-E_F}{2T}\right)}dE_F
\ee
and similarly for $\sigma_{\alpha\beta\gamma\delta}^{(3)}(\omega_1,\omega_2,\omega_3;\mu,T)$; here $\mu$ is the chemical potential at $T\neq 0$.

\subsection{Linear polarization, contribution IV\label{sec:lin-polariz}}

\begin{figure*}
\includegraphics[width=0.98\columnwidth]{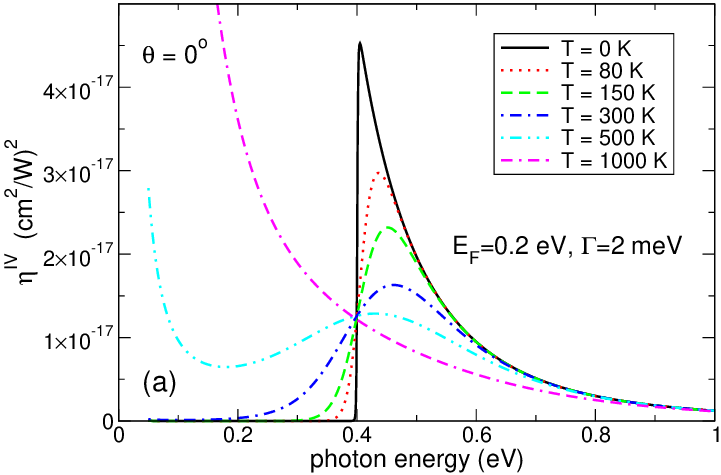}
\includegraphics[width=0.98\columnwidth]{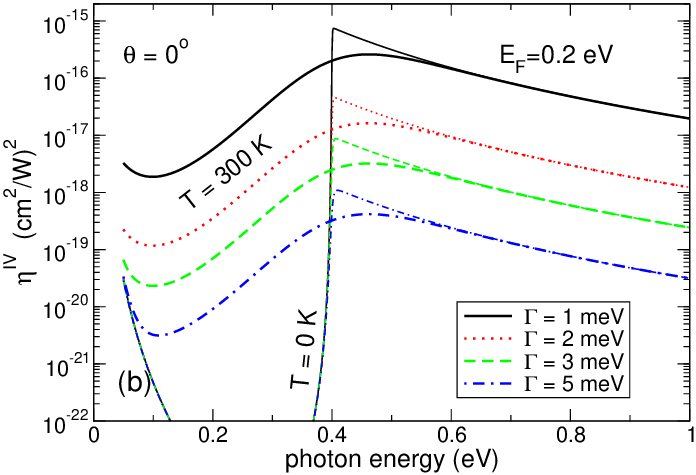}
\includegraphics[width=0.98\columnwidth]{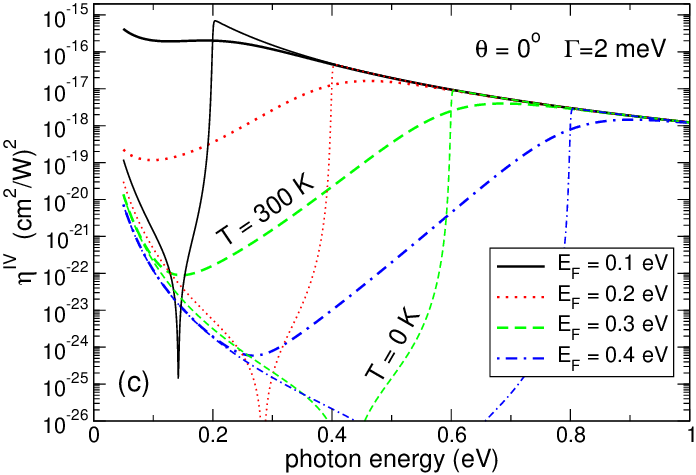}
\includegraphics[width=0.98\columnwidth]{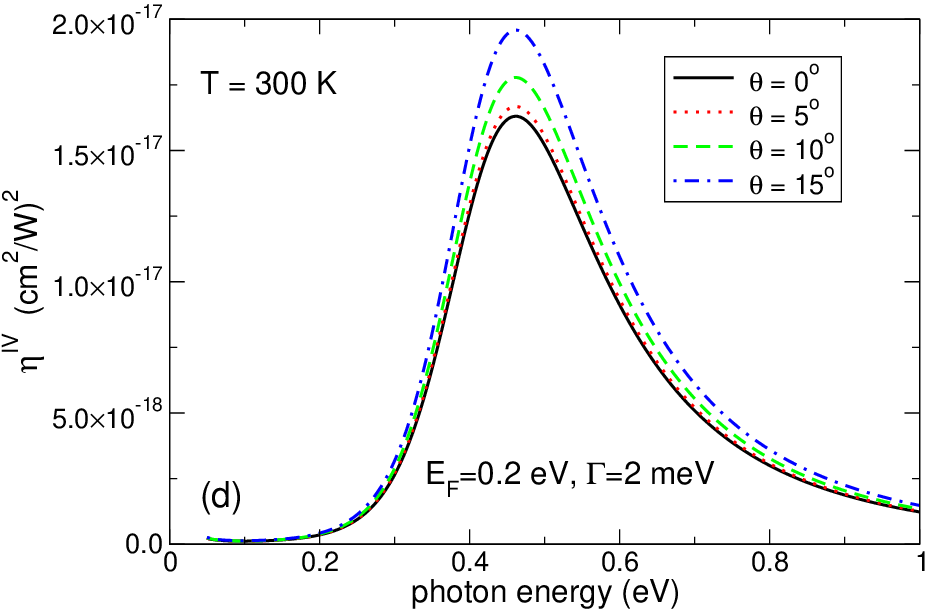}
\caption{\label{fig:etaIV} The function $\eta^{\rm IV}(\omega)$ defined in Eq. (\ref{intensIV0}) at (a) $E_F=0.2$ eV, $\Gamma=\hbar\gamma=2$ meV, $\theta=0^\circ$ and different temperatures, (b) $E_F=0.2$ eV, $\theta=0^\circ$, $T=0$ K (thin curves), and $T=300$ K (thick curves), and different values of $\Gamma$, (c) $\Gamma=2$ meV, $\theta=0^\circ$, $T=0$ K (thin curves), and $T=300$ K (thick curves), and different values of Fermi energy, (d) $E_F=0.2$ eV, $\Gamma=2$ meV, $T=300$ K, and different ellipticity parameters $\theta$.}
\end{figure*}

First we analyze the different contributions to the output wave intensity (\ref{intensII}) -- (\ref{intensVI}) at the ellipticity parameter $\theta=0$, i.e., when the incident probe ($S$) wave is linearly polarized. In this case all terms except one disappear and we have for the contribution IV (at $\theta=0$): 
\ba 
&&I_{\rm detect}^{\rm IV}\equiv \eta^{\rm IV}(\omega) I_SI_P^2
\nonumber \\ &=& \frac {36\pi^4}{c^4}I_SI_P^2
 \frac{\left|
\sigma^{(3)}_{xxxx}(\omega,\omega,-\omega)   -
\sigma^{(3)}_{xxyy}(\omega,\omega,-\omega) 
\right|^2}{\left|1+\frac{2\pi\sigma^{(1)}_{\omega}}{c}\right|^4 \left|1+\frac{2\pi\sigma^{(1)}_{\omega}} {c\cos\beta}\right|^4}.\nonumber \\
\label{intensIV0}
\ea
Figures \ref{fig:etaIV}(a)--\ref{fig:etaIV}(c) illustrate the frequency, Fermi energy, relaxation rate and temperature dependencies of the efficiency parameter $\eta^{\rm IV}(\omega)$ defined by the first equality in Eq. (\ref{intensIV0}). At low temperatures it is very small at $\hbar\omega\lesssim 2E_F$, has a sharp peak at $\hbar\omega= 2E_F$ and then decreases with frequency. When temperature grows [Fig. \ref{fig:etaIV}(a)], the peak becomes smoother and broader and the response function $\eta^{\rm IV}(\omega)$ strongly increases in the low-frequency regime $\hbar\omega\lesssim 2E_F$. At higher frequencies, $\hbar\omega\gg 2E_F$, in particular at the telecommunication wavelength $\lambda\simeq 1550 - 1600$ nm ($\hbar\omega\simeq 0.75$ eV) used in \cite{Dremetsika16}, the function $\eta^{\rm IV}(\omega)$ weakly depends on temperature and Fermi energy [Figs. \ref{fig:etaIV}(a) and \ref{fig:etaIV}(c)] but is very strongly influenced by the scattering rate parameter $\Gamma$ [Fig. \ref{fig:etaIV}(b)]. When $\Gamma$ changes from 5 meV down to 1 meV the value of $\eta^{\rm IV}(\omega)$ increases (at $\hbar\omega\simeq 0.75$ eV) by almost 3 orders of magnitude, from $\sim 10^{-19}$ (cm$^2$/W)$^2$ up to $\sim 6.2\times 10^{-17}$ (cm$^2$/W)$^2$. The absolute value of the intensity of the contribution IV is rather high. If we assume that $I_P\simeq 2\times 10^8$ W/cm$^2$, $I_S\simeq 10^7$ W/cm$^2$ (typical values in Ref. \cite{Dremetsika16}), and $\eta^{\rm IV}(\omega)\simeq 4\times 10^{-18}$ (cm$^2$/W)$^2$ (corresponding to $\hbar\omega\simeq 0.75$ eV and $\Gamma=2$ meV), we obtain $I_{\rm detect}^{\rm IV}\simeq 1.6\times 10^6$ W/cm$^2$, i.e., about 0.8\% of the pump power density and $\simeq 16$ \% of the probe signal power density.

\subsection{Elliptic polarization, contributions II -- VI\label{sec:ellip-polariz}}

\begin{figure*}
\includegraphics[width=0.98\columnwidth]{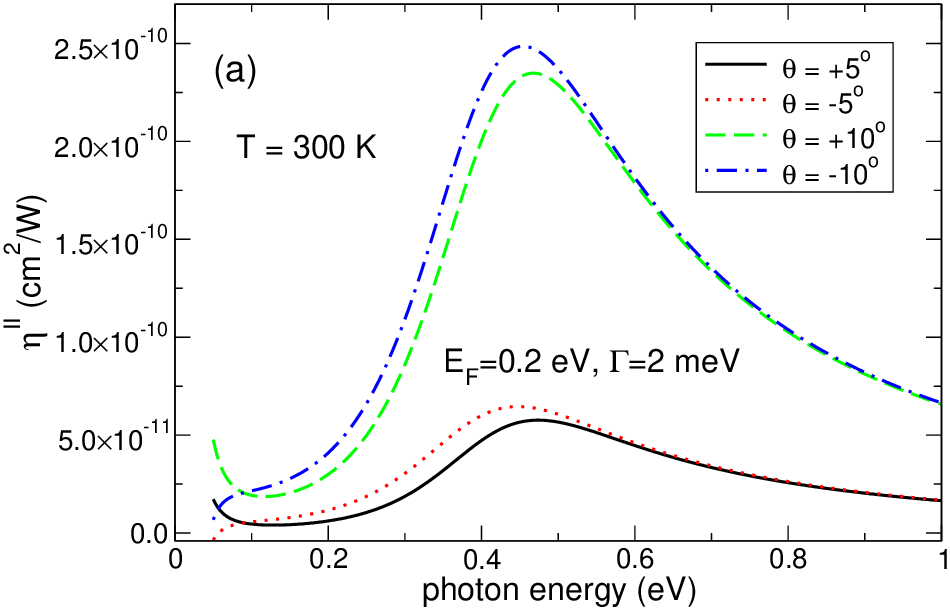}
\includegraphics[width=0.98\columnwidth]{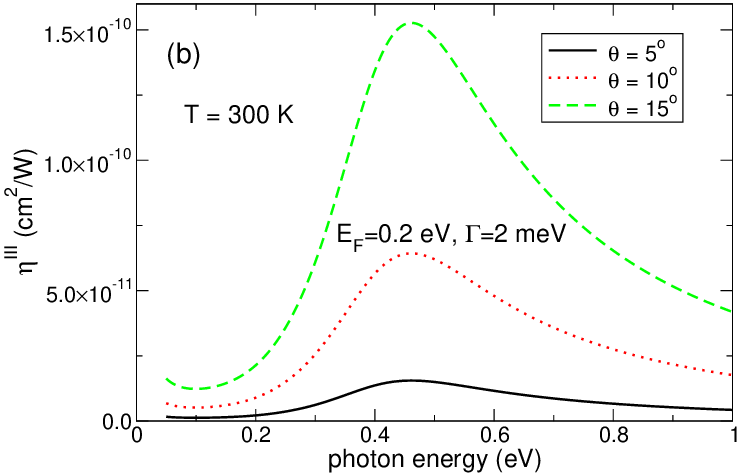}
\caption{\label{fig:etaII-III} The functions (a) $\eta^{\rm II}$ and (b) $\eta^{\rm III}$, defined in Eq. (\ref{etaII-III}) at $E_F=0.2$ eV, $\hbar\gamma=2$ meV, $T=300$ K, and different values of the ellipticity parameter $\theta$.
}
\end{figure*}

Now we consider the contributions II -- VI at a finite ellipticity $\theta\neq 0$. Figure \ref{fig:etaIV}(d) illustrates the growth of $\eta^{\rm IV}(\omega)$ with $\theta$. Notice that this growth is faster than linear; this will be additionally discussed below. Figures \ref{fig:etaII-III}(a) and \ref{fig:etaII-III}(b) exhibit the second-order response functions 
\be 
\eta^{\rm II}(\omega,\theta)\equiv \frac{I_{\rm detect}^{\rm II}}{I_SI_P}\ \ \textrm{ and } \ \ \eta^{\rm III}(\omega,\theta)\equiv \frac{I_{\rm detect}^{\rm III}}{I_S^2},\label{etaII-III}
\ee
defined using Eqs. (\ref{intensII}) and (\ref{intensIII}), at $E_F=0.2$ eV, $\Gamma=2$ meV, $T=300$ K, and different values of the ellipticity parameter $\theta$. Both functions have a shape similar to each other and to $\eta^{\rm IV}$ at $\theta=0$, see Fig. \ref{fig:etaIV}(d). The absolute values of $\eta^{\rm II}$ and $\eta^{\rm III}$ are also quite close: the former function is only about 3.5 times larger than the latter (compare the values at $\theta=10^\circ$). At first glance this seems to be rather unexpected, since the formulas (\ref{intensII}) and (\ref{intensIII}) show that $\eta^{\rm II}$ is proportional to $\sin\theta$ while $\eta^{\rm III}$ to $\sin^2\theta$; therefore one had to expect $\eta^{\rm II}\gg\eta^{\rm III}$ at small $\theta$. Moreover, as seen from Eqs. (\ref{intensII}) and (\ref{intensIII}), $\eta^{\rm III}$ is an even function of $\theta$, while $\eta^{\rm II}$ should be odd or at least should contain an essential odd contribution. Figure \ref{fig:etaII-III}(a), however, shows very close curves for positive and negative $\theta$'s, especially at $\hbar\omega>2E_F$. How do we explain this weak sensitivity of $\eta^{\rm II}$ to the polarization sense of the probe wave? 

Let us take a closer look at the functions $\eta^{\rm II}$ and $\eta^{\rm III}$. At large frequencies, $\hbar\omega\gg 2E_F$, the factor $2\pi\sigma_\omega^{(1)}/c$ is real and small as compared to unity, $2\pi\sigma_\omega^{(1)}/c=\pi\alpha/2\approx 0.011$, where $\alpha\approx 1/137$ is the fine-structure constant. The frequency dependencies of $\eta^{\rm II}$ and $\eta^{\rm III}$ are therefore mainly determined by Im ${\cal U}(\omega,\theta)$ and Re ${\cal V}(\omega,\theta)$ respectively, where ${\cal U}(\omega,\theta)$ and ${\cal V}(\omega,\theta)$ are defined in Eqs. (\ref{funcU}) and (\ref{funcV}). For these functions we have
\ba &&
\textrm{Im }{\cal U}(\omega,\theta)=
\nonumber \\ &-&\cos\theta\ \textrm{Im}\left[
\sigma^{(3)}_{xxxx}(\omega,\omega,-\omega)   -
\sigma^{(3)}_{xxyy}(\omega,\omega,-\omega) \right] \nonumber \\ &-&\sin\theta\ \textrm{Re}\left[
\sigma^{(3)}_{xxxx}(\omega,\omega,-\omega)   +
\sigma^{(3)}_{xxyy}(\omega,\omega,-\omega) \right],
\label{ImfuncU}
\ea
\ba
&&\textrm{Re }{\cal V}(\omega,\theta)=\cos 2\theta \ \textrm{Re} \left[\sigma^{(3)}_{xxxx}(\omega,\omega,-\omega)\right] 
  \nonumber \\ &-&
 2\left( 1+\cos 2\theta 
 \right) \ \textrm{Re}\left[\sigma^{(3)}_{xxyy}(\omega,\omega,-\omega)\right].
\label{RefuncV}
\ea 

\begin{figure*}
\includegraphics[width=0.98\columnwidth]{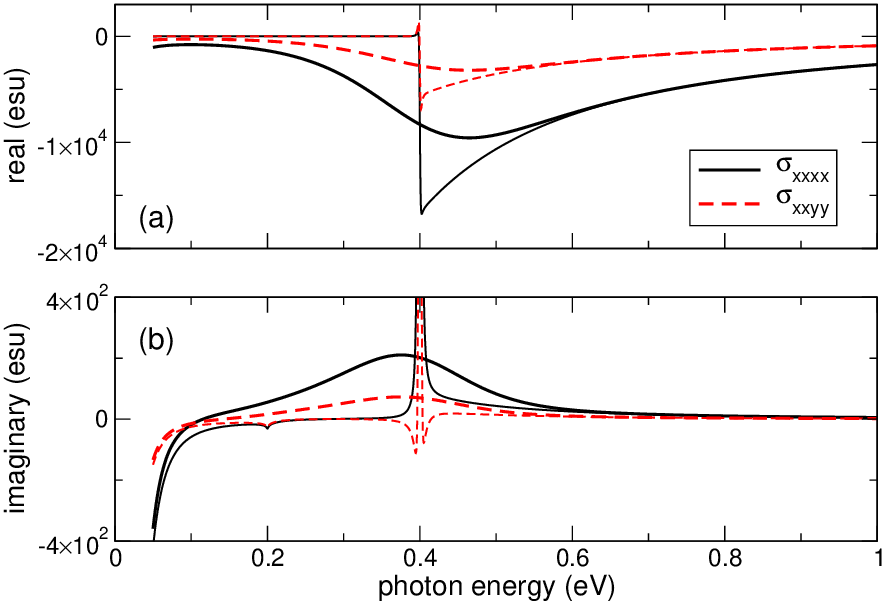}
\includegraphics[width=0.98\columnwidth]{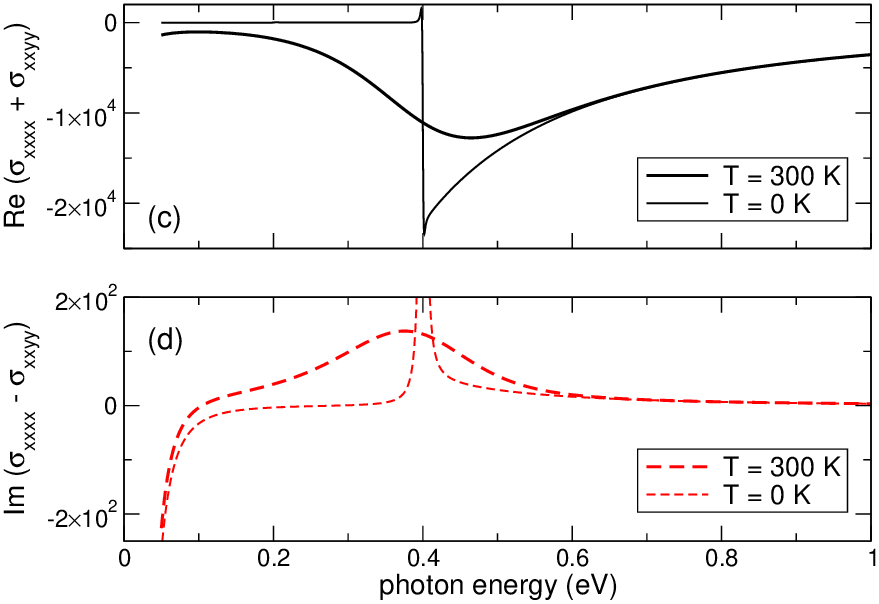}
\caption{\label{fig:Sxxyy} (a) The real and (b) imaginary parts of $\sigma^{(3)}_{xxxx}(\omega,\omega,-\omega)$ and $\sigma^{(3)}_{xxyy}(\omega,\omega,-\omega)$ at $E_F=0.2$ eV, $\hbar\gamma=2$ meV, and at room (thick curves) and zero (thin curves) temperatures. (c) The real part of $\left(
\sigma^{(3)}_{xxxx}(\omega,\omega,-\omega)   +
\sigma^{(3)}_{xxyy}(\omega,\omega,-\omega) \right)$ and (d) imaginary part of $\left(
\sigma^{(3)}_{xxxx}(\omega,\omega,-\omega)   -
\sigma^{(3)}_{xxyy}(\omega,\omega,-\omega) \right)$ at the same values of $E_F$ and $\Gamma$ and at room (thick curves) and zero (thin curves) temperatures.
}
\end{figure*}

Figures \ref{fig:Sxxyy}(a) and \ref{fig:Sxxyy}(b) show the real and imaginary parts of the functions $\sigma^{(3)}_{xxxx}(\omega,\omega,-\omega)$ and $\sigma^{(3)}_{xxyy}(\omega,\omega,-\omega)$ which enter Re ${\cal V}(\omega,\theta)$, Eq. (\ref{RefuncV}). At high frequencies $\hbar\omega>2E_F$ the imaginary parts of these functions are negligibly small as compared to their real parts, and their real parts differ approximately by a factor of 3. Similarly, Figs. \ref{fig:Sxxyy}(c) and \ref{fig:Sxxyy}(d) exhibit the real part of the sum $\left(
\sigma^{(3)}_{xxxx}(\omega,\omega,-\omega)   +
\sigma^{(3)}_{xxyy}(\omega,\omega,-\omega) \right)$ and the imaginary part of the difference $\left(
\sigma^{(3)}_{xxxx}(\omega,\omega,-\omega)   -
\sigma^{(3)}_{xxyy}(\omega,\omega,-\omega) \right)$ which determine Im ${\cal U}(\omega,\theta)$, see Eq. (\ref{ImfuncU}). Again, at high frequencies the real part of the sum is \textit{orders of magnitude} larger than the imaginary part of the difference. That is, the term in Im ${\cal U}(\omega,\theta)$ proportional to $\cos\theta$ is negligibly small as compared to the term proportional to $\sin\theta$, even at $\theta\simeq 1-5^\circ$. Since $\eta^{\rm II}$ is proportional to $\sin\theta[{\cal U}(\omega,\theta)]''$, this explains the weak sensitivity of $\eta^{\rm II}$ to the sign of $\theta$ and a similar order of magnitude of $\eta^{\rm II}$ and $\eta^{\rm III}$. This also explains the faster than linear (approximately quadratic) $\theta$ dependence of $\eta^{\rm IV}(\omega,\theta)\simeq |{\cal U}(\omega,\theta)|^2$ shown in Fig. \ref{fig:etaIV}(d).

Let us compare the absolute values of the intensities of the contributions II, III with that of the contribution IV. Assume again that $I_P\simeq 2\times 10^8$ W/cm$^2$, $I_S\simeq 10^7$ W/cm$^2$, and take for $\eta^{\rm II,III}(\omega)$ the values corresponding to $\hbar\omega\simeq 0.75$ eV, $E_F=0.2$ eV, $\Gamma=2$ meV, $T=300$ K, and $\theta=5^\circ$. We obtain $\eta^{\rm II}\simeq 3\times 10^{-11}$ cm$^2$/W and $\eta^{\rm III}\simeq 7.5\times 10^{-12}$ cm$^2$/W. This gives $I_{\rm detect}^{\rm II}\simeq 6\times 10^4$ W/cm$^2$ and $I_{\rm detect}^{\rm III}\simeq 7.5\times 10^2$ W/cm$^2$, as compared to $I_{\rm detect}^{\rm IV}\simeq 1.6\times 10^6$ W/cm$^2$ estimated in Section \ref{sec:lin-polariz}. These numbers are still sufficiently high to be experimentally observed, but the ``second-order'' terms II and III turn out to be smaller than the ``third-order'' term IV under the same conditions.

\begin{figure*}
\includegraphics[width=0.98\columnwidth]{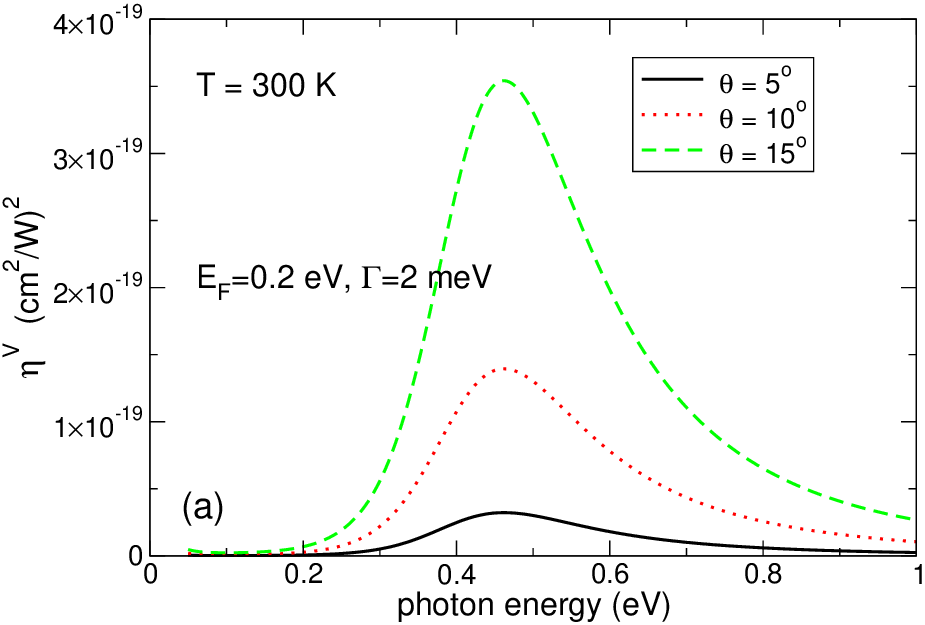}
\includegraphics[width=0.98\columnwidth]{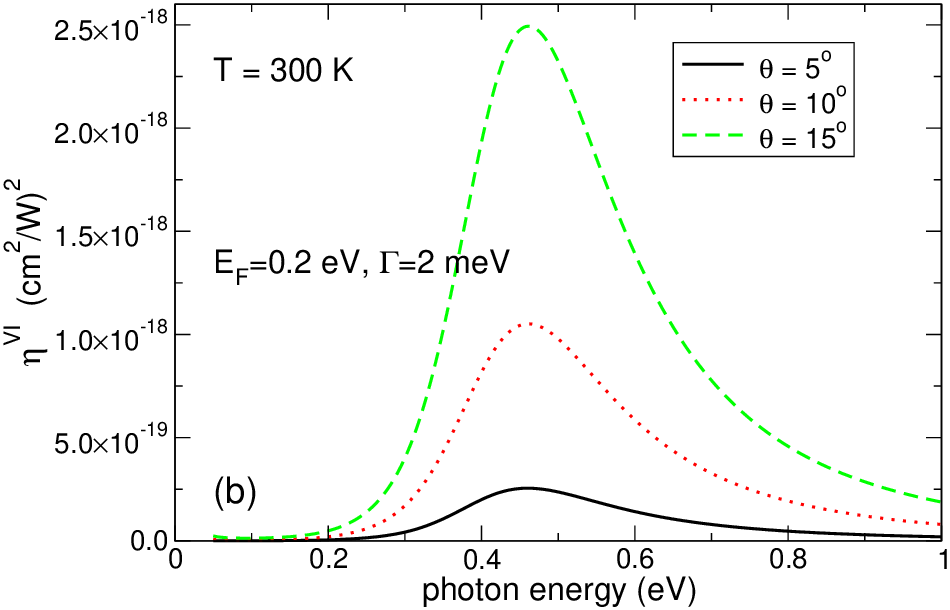}
\caption{\label{fig:etaV-VI} The functions (a) $\eta^{\rm V}$ and (b) $\eta^{\rm VI}$ defined in Eq. (\ref{def-etaVVI}), at $E_F=0.2$ eV, $\hbar\gamma=2$ meV, $T=300$ K and different values of the ellipticity parameter $\theta$.
}
\end{figure*}

Figures \ref{fig:etaV-VI}(a) and \ref{fig:etaV-VI}(b) exhibit the third order response functions 
\be 
\eta^{\rm V}(\omega,\theta)\equiv \frac{I_{\rm detect}^{\rm V}}{I_S^3}\ \ \textrm{ and } \ \ \eta^{\rm VI}(\omega,\theta)\equiv \frac{I_{\rm detect}^{\rm VI}}{I_S^2I_P},\label{def-etaVVI}
\ee
defined according to Eqs. (\ref{intensV}) and (\ref{intensVI}). Their frequency dependencies are similar to other contributions. Their absolute values at $\hbar\omega\simeq 0.75$ eV, $E_F=0.2$ eV, $\Gamma=2$ meV, $T=300$ K and $\theta=5^\circ$ are $\eta^{\rm V}(\omega,\theta)\simeq 7.8\times 10^{-21}$ (cm$^2$/W)$^2$ and $\eta^{\rm VI}(\omega,\theta)\simeq 6\times 10^{-20}$ (cm$^2$/W)$^2$, which gives (again at $I_P\simeq 2\times 10^8$ W/cm$^2$ and $I_S\simeq 10^7$ W/cm$^2$) $I_{\rm detect}^{\rm V}\simeq 7.8$ W/cm$^2$ and $I_{\rm detect}^{\rm VI}\simeq 1.2\times 10^3$ W/cm$^2$. 

In Table \ref{tab:tableEstim} we summarize the estimates obtained for different contributions to $I_{\rm detect}$ on the basis of the $\sigma^{(3)}$ model of Refs. \cite{Cheng15,Mikhailov16a}. The contribution IV remains the largest, followed by the terms II and VI. The smallest contribution is V since it is proportional to $I_S^3$. By varying the input intensities, e.g., making $I_S$ stronger than $I_P$, one could modify the mutual relations between the terms II--VI. Notice that the calculated intensities of all five nonlinear contributions II--VI, Table \ref{tab:tableEstim}, remain much smaller than the intensity of the incident pump wave. This shows that the third-order response theory described by the tensor $\sigma^{(3)}_{\alpha\beta\gamma\delta}(\omega_1,\omega_2,\omega_3)$ is still valid at the relatively large intensity of radiation $\sim 200$ MW/cm$^2$ which was used in Ref. \cite{Dremetsika16}. 

\begin{table}[t]
\caption{\label{tab:tableEstim}
Estimates of different contributions $I_{\rm detect}^{\rm II,III,IV,V,VI}$ (in W/cm$^2$) to the electromagnetic wave intensity registered by the detector at $\hbar\omega\simeq 0.75$ eV, $E_F=0.2$ eV, $\Gamma=2$ meV, $\theta=5^\circ$, $I_P\simeq 2\times 10^8$ W/cm$^2$, and $I_S\simeq 10^7$ W/cm$^2$. }
\begin{ruledtabular}
\begin{tabular}{ccccc}
II & III & IV & V & VI \\
\colrule
$6\times 10^4$ & $7.5\times 10^2$ & $1.6\times 10^6$ & $7.8$ & $1.2\times 10^3$ \\
\end{tabular}
\end{ruledtabular}
\end{table}

\subsection{Comparison with experiment\label{sec:comparison}} 

\begin{figure}
\includegraphics[width=\columnwidth]{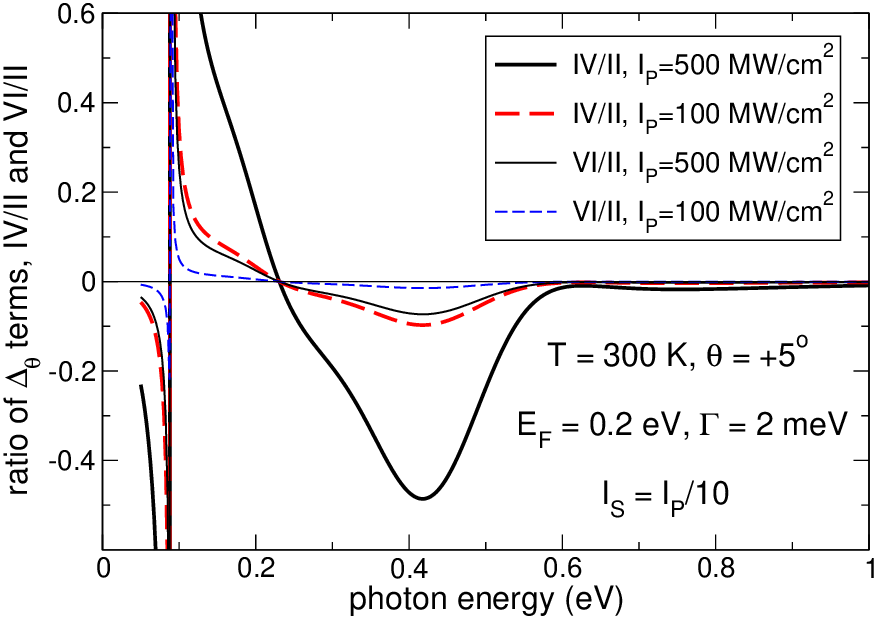}
\caption{\label{fig:Deltas} The ratio of the intensities $\left[\Delta_\theta I_{\rm detect}^{f_S+ f_P}(\omega,\theta)\right]_{\rm IV}/\left[\Delta_\theta I_{\rm detect}^{f_S+ f_P}(\omega,\theta)\right]_{\rm II}$ and $\left[\Delta_\theta I_{\rm detect}^{f_S+ f_P}(\omega,\theta)\right]_{\rm VI}/\left[\Delta_\theta I_{\rm detect}^{f_S+ f_P}(\omega,\theta)\right]_{\rm II}$, defined in Eqs. (\ref{DiffThetaII}) and (\ref{DiffThetaIV+VI}), at $E_F=0.2$ eV, $\hbar\gamma=2$ meV, $T=300$ K, $\theta=5^\circ$ and $I_P/I_S=10$ as a function of radiation frequency. The incident pump wave intensity $I_P=500$ MW/cm$^2$ and $I_P=100$ MW/cm$^2$. For the modulations depths in (\ref{DiffThetaII}) and (\ref{DiffThetaIV+VI}) the numbers $\alpha_P=\alpha_S=1$ are taken.
}
\end{figure}

Now consider the quantities that have been measured in Ref. \cite{Dremetsika16} and discussed in Section \ref{sec:whatwas}, Eqs. (\ref{DiffTheta}) -- (\ref{DiffThetaIV+VI}), and quantitatively compare our results with the experimental data. First, we check under which conditions the contributions IV and VI, Eq. (\ref{DiffThetaIV+VI}), are small as compared to the term II, Eq. (\ref{DiffThetaII}). Figure \ref{fig:Deltas} shows the ratios IV/II and VI/II as a function of frequency at a typical set of input parameters. 
One sees that at $\hbar\omega\gtrsim 0.6$ eV ($\hbar\omega\gtrsim 3E_F$) the contributions IV and VI are small indeed as compared to II (in the experiment \cite{Dremetsika16} $\hbar\omega\simeq 0.75$ eV). Around the photon energy $\hbar\omega\simeq 0.4$ eV, which corresponds to the interband resonance at $\hbar\omega=2E_F$, and at $I_P\simeq 500$ MW/cm$^2$ (the power density used in Ref. \cite{Dremetsika16}) the contributions IV and VI may achieve 50\% and 10\% of II respectively, but at lower intensities $I_P\lesssim 100$ MW/cm$^2$ they still can be neglected. At even lower photon energies ($\lesssim E_F=0.2$ eV) the contributions IV and VI become dominant. The analysis of Ref. \cite{Dremetsika16} which led to the simple expression (\ref{experDiff}), although valid at near-IR frequencies, would not be correct at frequencies $\hbar\omega\lesssim 2E_F$. 

\begin{figure*}
\includegraphics[width=0.98\columnwidth]{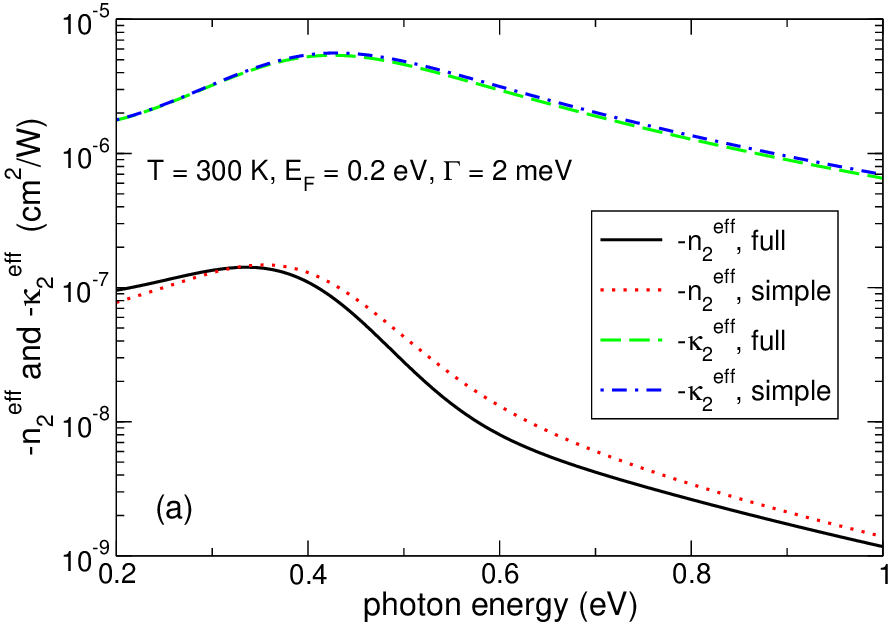}
\includegraphics[width=0.98\columnwidth]{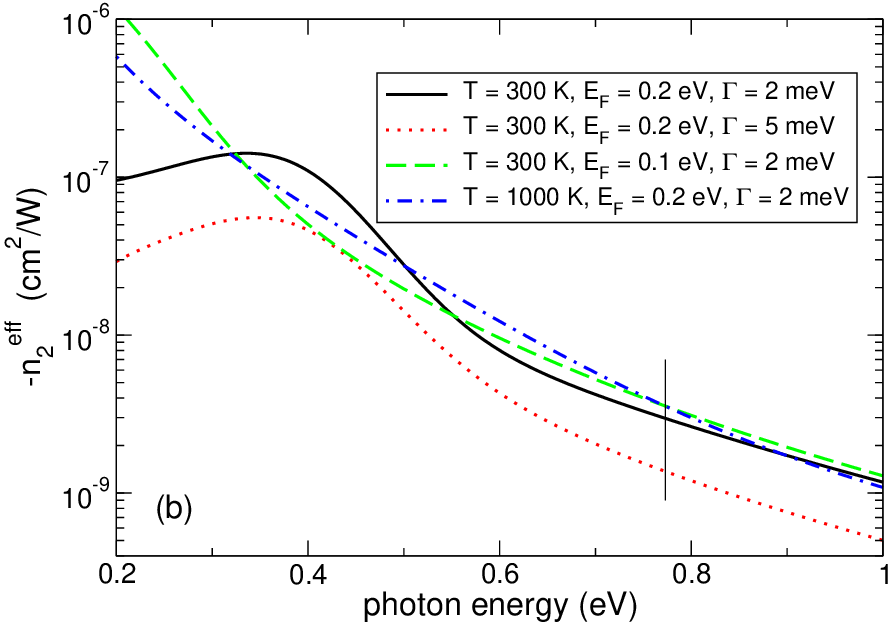}
\caption{\label{fig:N2eff} (a) The negative \textit{effective} nonlinear refractive index $-n_2^{\rm eff}$ and absorption coefficient $-\kappa_2^{\rm eff}$ calculated using the full formulas (\ref{correspondence2}), (\ref{correspondence3}) and the simplified one (\ref{correspondence2simple}), at $E_F=0.2$ eV, $T=300$ K and $\Gamma=\hbar\gamma=2$ meV. (b) The negative \textit{effective} nonlinear refractive index $-n_2^{\rm eff}$ (full formula) for a few different values of $E_F$, $T$ and $\hbar\gamma$. The thin vertical line corresponds to the wavelength $\lambda=1600$ nm ($\hbar\omega=0.773$ eV) used in Ref. \cite{Dremetsika16}.}
\end{figure*}

Now assume that the conditions under which the contributions IV and VI can be neglected are satisfied. Then the \textit{effective} nonlinear refractive index $n_2^{\rm eff}$ and the \textit{effective} nonlinear absorption coefficient $\kappa_2^{\rm eff}$ are determined by formulas (\ref{correspondence2}) and (\ref{correspondence3}) respectively. These formulas can be simplified further if we neglect the factors $\sim 2\pi\sigma_\omega^{(1)}/c$ in the denominators of Eqs. (\ref{correspondence2}) and (\ref{correspondence3}). Then one gets 
\ba 
&&n_2^{\rm eff} +i\kappa_2^{\rm eff} \nonumber \\ & \widetilde{\Leftrightarrow}& \ 
i\frac {6\pi^2}{\omega cL^{\rm eff}} 
\left[
\sigma^{(3)}_{xxxx}(\omega,\omega,-\omega) -
\sigma^{(3)}_{xxyy}(\omega,\omega,-\omega)\right].
\label{correspondence2simple}
\ea
Figure \ref{fig:N2eff}(a) shows the frequency dependence of $n_2^{\rm eff}$ and $\kappa_2^{\rm eff}$ at a typical set of input parameters. Both quantities are negative in the shown interval of photon energies, therefore plotted are  $-n_2^{\rm eff}$ and $-\kappa_2^{\rm eff}$, and exhibited are curves calculated according to Eqs. (\ref{correspondence2}) and (\ref{correspondence3}) (the full formulas) and according to Eq. (\ref{correspondence2simple}) (the simplified formula). The difference between the full and simplified formulas for $\kappa_2^{\rm eff}$ is small. For $n_2^{\rm eff}$ the difference is quite noticeable, for example, at $\hbar\omega\simeq 0.77$ eV the full and simplified formulas give $n_2^{\rm eff}\approx -3\times 10^{-9}$ cm$^2$/W and $n_2^{\rm eff}\approx -4\times 10^{-9}$ cm$^2$/W, respectively. 

The absolute value of \textit{effective} $\kappa_2^{\rm eff}$ is several orders of magnitude larger than that of $n_2^{\rm eff}$, e.g. $|\kappa_2^{\rm eff}|\approx 1.43\times 10^{-6}$ cm$^2$/W and $|n_2^{\rm eff}|\approx 3\times 10^{-9}$ cm$^2$/W at 0.77 eV, $|\kappa_2^{\rm eff}|/|n_2^{\rm eff}|\approx 477$. At lower frequencies corresponding to the interband resonance at $\hbar\omega\simeq 2E_F=0.4$ eV the absolute value of $n_2^{\rm eff}$ is 2 orders of magnitude larger than at the telecommunication frequency $\sim 0.77$ eV. (One should remember, however, that around the interband resonance the terms IV and VI may become essential and should in general be taken into account.) The negative sign of $\kappa_2^{\rm eff}$ corresponds to the absorption saturation which was experimentally observed in graphene \cite{Bao09,Popa10,Popa11,Bianchi17} and topological insulators \cite{Chen15}. The negative sign of $n_2^{\rm eff}$ implies a self-defocusing nonlinearity and was observed in Refs. \cite{Dremetsika16,Vermeulen16}. 

In Ref. \cite{Dremetsika16} the value of $n_2^{\rm eff}=-1.1\times 10^{-9}$ cm$^2$/W was extracted from the OHD-OKE experiment at $\lambda=1600$ nm for monolayer graphene. The quantity $n_2^{\rm eff}\approx -3\times 10^{-9}$ cm$^2$/W which one gets from Fig. \ref{fig:N2eff}(a) at $E_F=0.2$ eV, $T=300$ K and $\Gamma=2$ meV is a bit larger in absolute value. In Fig. \ref{fig:N2eff}(b) we plot several theoretical curves for $n_2^{\rm eff}$ varying the Fermi energy, temperature, and the relaxation rate. One sees that changing the Fermi energy (green dashed curve) or temperature (blue dot-dashed curve) does not influence this number substantially, in accordance with our discussion in Sec. \ref{sec:lin-polariz}. But by slightly changing the effective relaxation rate $\Gamma\to 5$ meV one can get better agreement with the experiment, $n_2^{\rm eff}\to -1.37\times 10^{-9}$ cm$^2$/W. Thus the theory and experiment \cite{Dremetsika16} agree quite well with each other, both in terms of the sign and the absolute values of the measured nonlinear parameters of graphene.

The value of $n_2^{\rm eff}$ extracted from the $Z$-scan measurement in Ref. \cite{Dremetsika16} ($n_2^{\rm eff}=-2\times 10^{-8}$ cm$^2$/W) quite substantially differed from the one found from the OHD-OKE measurements. As we mentioned above (Sec. \ref{sec:whatwas}), the \textit{effective} $n_2^{\rm eff}$ and $\kappa_2^{\rm eff}$ extracted from the $Z$-scan experiment are essentially \textit{different quantities} since they are determined by a different combination of the third conductivity tensor components. This confirms, once again, the inappropriateness of using essentially 3D quantities $n_2$ and $\kappa_2$ for a description of 2D crystals. A further discussion of the $Z$-scan technique for graphene is beyond the scope of this paper. 

\section{Summary and conclusions\label{sec:conclus}} 

We have presented a detailed theoretical analysis of the OHD-OKE technique of measuring nonlinear properties of graphene and other 2D materials. Let us summarize the results obtained. In Sec. \ref{sec:theory} we have derived analytical formulas (\ref{intensII}) -- (\ref{intensVI}) which allow one to experimentally study all components of the third-order conductivity tensor measuring the intensity of five different nonlinear output signal contributions to the OHD-OKE signal. These formulas are not related to any model of $\sigma^{(3)}_{\alpha\beta\gamma\delta}$ and can be used for analysis of the nonlinear response of any material which is much thinner than the radiation wavelength, especially of 2D materials with one or a few atomic layers. 

In Sec. \ref{sec:theory-model} we have specified our general results, having used the model of $\sigma^{(3)}_{\alpha\beta\gamma\delta}$ developed in Refs. \cite{Cheng15,Mikhailov16a} and studied the frequency, doping, temperature, and relaxation rate dependencies of different contributions to the output signal. We have predicted a rich behavior of the Kerr response as a function of all these factors, including a rather strong growth of the effect near the interband resonance transition at $\hbar\omega\simeq 2E_F$ as well as in the low-frequency region $\hbar\omega\lesssim E_F$. We have compared our results with the experimental findings of Dremetsika et al. \cite{Dremetsika16} and found good agreement with their data. 

All results of this paper have been obtained for an isolated graphene layer. In experiments graphene typically lies on a dielectric substrate. If the substrate thickness is small as compared to the radiation wavelength our results remain quantitatively valid. If graphene lies on a substrate with a rough backside which does not reflect radiation the denominators $\left[1+2\pi\sigma_\omega^{(1)}/c\right]$ in the above-derived formulas should be replaced by $\left[(n+1)/2+2\pi\sigma_\omega^{(1)}/c\right]$, where $n$ is the (linear) refractive index of the substrate. The analysis of more complex cases is beyond the scope of the present work. The influence of different types of the substrate resonances (Fabry-P\'erot, optical phonon resonances) on the third harmonic generation effect was comprehensively studied in Refs. \cite{Savostianova15,Savostianova17a}; the role of these resonances in the Kerr response can be understood similarly.

Most of experiments on the nonlinear optical response of graphene have been performed so far at a single or a few frequencies, single or a few values of the Fermi energy and at unknown or uncontrolled values of the effective relaxation rate. The theory predicts very interesting dependencies of the nonlinear graphene response on all these parameters. Therefore further extensive experimental studies of Kerr and other nonlinear effects in graphene are highly desirable, promising important fundamental discoveries and useful optoelectronic and photonic applications. 

\begin{acknowledgments}
We are grateful to Evdokia Dremetsika for many useful discussions concerning details of the experiment \cite{Dremetsika16}. The work has received funding from the European Union's Horizon 2020 research and innovation programmes Graphene Core 1 and Graphene Core 2 under Grant Agreements No. 696656 and No. 785219.
\end{acknowledgments}

%


\end{document}